\documentclass[final,5p,times,authoryear]{elsarticle}
\usepackage[utf8]{inputenc} 
\usepackage[T1]{fontenc}
\usepackage{layout}
\usepackage{multirow}
\usepackage{booktabs}
\usepackage[utf8]{inputenc}
\usepackage{comment}
\usepackage{lineno}
\nolinenumbers
\newcommand{\sampleline}[2][]{%
    \tikz[baseline=-0.5ex] \draw[#1, line width=#2] (0,0) -- (1.5,0);
}

\usepackage{epsfig}
\usepackage{amssymb}
\usepackage{amsmath}
\usepackage{graphicx}
\usepackage[labelformat=simple]{subcaption}
\usepackage{multicol}
\usepackage{soul}
\usepackage[dvipsnames]{xcolor}
\usepackage{tabularx}
\usepackage{mathtools,tikz}
\usepackage{placeins}
\usepackage{algorithm2e}
\usepackage[normalem]{ulem}
\SetKwInput{KwModel}{Model}
\makeatletter

\usepackage[dvipsnames]{xcolor}
\definecolor{myyellow}{HTML}{FFF2CC}
\definecolor{mygreen}{HTML}{D5E8D4}
\definecolor{mylinear}{HTML}{DAE8FC}
\definecolor{myblue}{HTML}{9999FF}
\definecolor{darkgreen}{HTML}{006400}
\usepackage[pdfencoding=auto, psdextra, colorlinks = true]{hyperref}
\hypersetup{
    colorlinks=true,
  citecolor=violet,
  linkcolor=red,
  urlcolor=green,
  }
\usepackage{amsthm}

\journal{International Journal of Heat and Fluid Flow}

\begin{document}

\begin{frontmatter}



\title{\textbf{Upstream history quantification and scale-decomposed energy analysis for weak-to-strong adverse-pressure-gradient turbulent boundary layers}}


\author[inst1]{Atharva Mahajan}

\author[inst2]{Rahul Deshpande\corref{cor1}}
\ead{raadeshpande@gmail.com}

\author[inst3]{Taygun R. Gungor}
\author[inst3]{Yvan Maciel}

\author[inst4]{Ricardo Vinuesa\corref{cor1}}
\ead{rvinuesa@kth.se}

\cortext[cor1]{Corresponding author.}

\affiliation[inst1]{organization={Department of Mechanical Engineering},
            addressline={BITS Pilani}, 
            city={Hyderabad},
            postcode={500078},
            country={India}}
\affiliation[inst2]{organization={Department of Mechanical Engineering},
            addressline={University of Melbourne}, 
            city={Parkville},
            postcode={VIC 3010},
            country={Australia}}
\affiliation[inst3]{organization={Department of Mechanical Engineering},
            addressline={Laval University},
            city={Quebec City},
            postcode={QC},
            country={G1V 0A6 Canada}}
\affiliation[inst4]{organization={FLOW, Engineering Mechanics},
            addressline={KTH Royal Institute of Technology}, 
            postcode={SE-100 44},
            city={Stockholm},
            country={Sweden}}

\begin{abstract}

The present study delineates the effects of pressure gradient history and local disequilibration on the small and large-scale energy in turbulent boundary layers (TBLs) imposed with a broad range of adverse-pressure-gradients (APG). 
This is made possible by analyzing four published high-fidelity APG TBL databases, which span weak to strong APGs and cover dynamic conditions ranging from near-equilibrium to strong disequilibrium. 
These databases enable the development of a methodology to understand the effects of PG history and local disequilibration, the latter defined here as the local streamwise rate of change of the pressure force contribution in the force balance.
The influence of PG history on TBL statistics is quantified by the accumulated PG parameter ($\overline{\beta}$), proposed previously by \citet{vinuesaRevisitingHistoryEffects2017} to study integral quanitites, which is compared here between cases at matched local PG strength ($\beta$), Reynolds number ($Re$) and ${\rm d}{\beta}/{{\rm d}{Re}}$ at nominally similar orders of magnitude. Here, $\beta$ denotes a general umbrella term used for pressure gradient parameters that is estimated using different scaling parameters in this study.
While the effects of local disequilibration (${\rm d}{\beta}/{{\rm d}{Re}}$) are investigated by considering TBL cases at matched $\beta$, $Re$, and fairly matched $\overline{\beta}$.
This enables analysis of accumulated PG history and local disequilibration effects separately, where applicable, to highlight qualitative differences in statistical trends.
It is found that $\overline{\beta}$ cannot unambiguously capture history effects when ${\rm d}{\beta}/{{\rm d}{Re}}$ levels are significantly high, as it does not account for the delayed response of the mean flow and turbulence, nor the attenuation of the pressure gradient effect with distance. 
In two comparisons of APG TBLs under strong non-equilibrium, the values of $\overline{\beta}$ and ${\rm d}{\beta}/{\rm d}{Re}$ expressed using Zagarola-Smits scaling were found to be consistent with the trends in mean velocity defect and Reynolds stresses noted previously for weak APG TBLs.
While an increase in $\overline{\beta}$ is associated with energisation of both the small and large scales in the outer regions of APG TBLs, it affects only the large scales in the near-wall region.
This confirms the ability of near-wall small scales to rapidly adjust to changes in PG strength. 
By attempting to provide a structured parametric methodology to isolate effects of PG history and local disequilibration, this study reports the influence of these effects on turbulent flow statistics across the widest APG strengths documented in the literature.

\end{abstract}


\end{frontmatter}
%

\section{Introduction}

\label{sec:intro}

Adverse-pressure-gradient (APG) turbulent boundary layers (TBLs) are encountered in various settings such as the flow over wings, automobiles and in turbomachinery applications. 
Hence, understanding the effects of pressure gradients on TBLs is crucial for developing flow control and drag reduction strategies for improving efficiencies of these engineering applications. 
Equilibrium TBLs represent the simplest but rarest forms of PG TBLs, characterized by a constant force balance as the flow develops \citep{rotta1950theorie, clauserTurbulentBoundaryLayer1956a, townsendPropertiesEquilibriumBoundary1956}. 
Here, equilibrium refers to the unique state of the TBL where the flow properties remain dynamically similar despite the flow developing/evolving in the streamwise direction \citep{maciel2006self,devenportEquilibriumNonequilibriumTurbulent2022}. 
Achieving equilibrium in TBLs is, however, challenging due to its complex multi-scale behavior, with each layer of the flow governed by characteristic length/time scales and force balances. 
The concept of equilibrium is inherently related with self-similarity; a flow is said to be self-similar if its statistical properties are solely influenced by local flow variables \citep{kitsiosDirectNumericalSimulation2017}. 
However, almost all the APG TBLs in practise exist in a non-equilibrium state, where turbulence statistics exhibit a multi-parameter dependence on local PG strength, flow Reynolds numbers, upstream PG/history as well as local disequilibration effects \citep{bobkeHistoryEffectsEquilibrium2017,gungor2024a}.
Historically, the first steps in understanding PG effects on TBL statistics were focused on quantification of the local PG strength, such as by the Clauser parameter $\beta_C$ \citep{clauserTurbulentBoundaryLayer1956a}, involving estimation of the ratio of pressure force to a reference force corresponding to the TBL.
Per definition, $\beta_{C} =$ $(\delta^*/\tau_w)(dP/dx)$, where $\delta^*$ is the TBL displacement thickness, $\tau_w$ is the mean wall shear stress and $(dP/dx)$ is the mean pressure-gradient over the streamwise distance x. 

\begin{table}[t!]
\centering
\footnotesize
\fbox{%
  \begin{minipage}{\columnwidth}
    \textbf{Nomenclature} \\[0.8em]
    \begin{tabularx}{\columnwidth}{p{2.4cm} X}
      \multicolumn{2}{l}{\textbf{Symbols}} \\
      $C_f$ & Skin friction coefficient \\
      $H_{12}$ & Shape factor, $\delta^*/\theta$ \\
      $l_\tau$ & Friction-viscous length scale, $\nu/U_\tau$ \\
      $Re_\tau$ & Friction Reynolds number, $U_\tau \delta/\nu$ \\
      $Re_\theta$ & Reynolds number based on $\theta$, $U_e \theta/\nu$ \\
      $Re_{ZS}$ & Zagarola-Smits Reynolds number \\
      $U$ & Mean streamwise velocity \\
      $U_e$ & Boundary layer edge velocity \\
      $U_\tau$ & Friction velocity \\
      $U_{ZS}$ & Zagarola-Smits velocity scale \\
      $V$ & Mean wall-normal velocity \\
      $W$ & Mean spanwise velocity \\
      $u, v, w$ & Velocity fluctuations in $x, y, z$ directions \\ [0.7em]
      $\overline{u^2}$, $\overline{v^2}$, $w^2$ & Reynolds normal stresses \\ [0.7em]
      $u^2_{LS}$ & Large-scale contribution to $\overline{u^2}$\\ [0.7em]
      $u^2_{SS}$ & Small-scale contribution to $\overline{u^2}$\\ [0.7em]
      $v^2_{LS}$ & Large-scale contribution to $\overline{v^2}$\\ [0.7em]
      $v^2_{SS}$ & Small-scale contribution to $\overline{v^2}$\\ [0.7em]
      $x$ & Streamwise coordinate \\
      $y$ & Wall-normal coordinate \\
      $z$ & Spanwise coordinate \\
      \\
      \multicolumn{2}{l}{\textbf{Greek Symbols}} \\
      $\beta_C$ & Clauser pressure-gradient parameter \\
      $\beta_{ZS}$ & Zagarola-Smits pressure-gradient parameter \\
      ${d\beta}/{dRe}$ & Local disequilibration parameter \\
      $\overline{\beta}$ & Accumulated pressure-gradient parameter \\
      $\delta$ & Boundary layer thickness \\
      $\delta^*$ & Displacement thickness \\
      $\lambda_{z,c}$ & Spanwise wavelength cutoff for scale decomposition \\
      $\nu$ & Kinematic viscosity \\
      $\phi_{uu}$ & Power spectral density of $u$ \\
      $\phi_{vv}$ & Power spectral density of $v$ \\
      $\rho$ & Fluid density \\
      $\tau_w$ & Wall shear stress \\
      $\theta$ & Momentum thickness \\
      \\
      \multicolumn{2}{l}{\textbf{Subscripts and Superscripts}} \\
      $()_0$ & Value under reference or neutral condition \\
      $\overline{()^2}$ & Variances \\
      $\overline{()}$ & Time or ensemble average \\
      \\
      \multicolumn{2}{l}{\textbf{Abbreviations}} \\
      APG & Adverse pressure gradient \\
      DNS & Direct numerical simulation \\
      LES & Large-eddy simulation \\
      TBL & Turbulent boundary layer \\
      ZPG & Zero pressure gradient \\
    \end{tabularx}
  \end{minipage}
}
\end{table}
Specifically, $\beta_C$ quantifies the relative strength of the streamwise pressure force to the wall shear force, providing a global measure of the force balance across the boundary layer.
Past studies have noted the variation in pressure gradient strength to lead to notable deviations in turbulent flow statistics from classical scaling theories, such as the log-law scaling of the mean streamwise velocity \citep{colesLawWakeTurbulent1956,marusicLogarithmicRegionWall2013,knoppExperimentalAnalysisLog2021}. 
Additionally, APG TBLs are characterized by greater mean velocity defects and increased intensities of Reynolds stress components, particularly in their outer regions \citep{skareTurbulentEquilibriumBoundary1994, harunPressureGradientEffects2013}. 
Notably, all the Reynolds stresses in APG TBLs display a distinctive outer peak at low friction Reynolds numbers, which, in ZPG TBLs are expected to appear only at very high $Re_{\tau}$ \citep{willertNearwallStatisticsTurbulent2017,deshpandeStreamwiseEnergytransferMechanisms2024}. 
Here the friction Reynolds number is defined as $Re_{\tau} = U_{\tau}\delta/\nu$, where $U_{\tau}$ is the mean friction velocity, $\delta$ is the mean boundary layer thickness, and $\nu$ is the kinematic viscosity. 
The association of these intensified Reynolds stresses with their corresponding Reynolds transport equations, for varying APG strengths, has been discussed previously by \cite{gungorEnergyTransferMechanisms2022} and \cite{deshpandeStreamwiseEnergytransferMechanisms2024}.
On the other hand, the effects of $Re_\tau$ on the outer-region energy of APG TBLs have been investigated by \citet{montyParametricStudyAdverse2011}, \citet{pozueloAdversepressuregradientTurbulentBoundary2022a} and
\cite{deshpandeReynoldsnumberEffectsOuter2023d}.
Specifically, \citet{deshpandeReynoldsnumberEffectsOuter2023d} employed an experimental setup that minimized the effects of history and disequilibration by controlling the airflow from the wind tunnel ceiling.
They also analyzed the small- and large-scale contributions to the Reynolds stresses in weak-to-moderately strong APGs, which unraveled the decrease in their small-scale energy for an increase in $Re_{\tau}$ \citep{deshpandeReynoldsnumberEffectsOuter2023d}. 
Notably, this $Re_{\tau}$-trend is in contrast with those observed for increasing APG magnitudes and upstream history effects, which past studies have held responsible for increasing both the small- and large-scale energies \citep{bobkeHistoryEffectsEquilibrium2017,montyParametricStudyAdverse2011,tanarroEffectAdversePressure2020a}.
However, it still remains unknown if PG history influences the small and large-scale energies in the same way for strong APG cases, which this study will investigate.

Flows in practical applications are characterized by unique distributions of streamwise variation rates of the PG (local disequilibration), which in turn leads to distinct `accumulated' PG histories. 
It is therefore crucial to independently understand the impact of both of these effects on local turbulent flow statistics.
The effect of local disequilibration can be explained as the streamwise change rate of the pressure force balance \citep{gungor2024a}.
\citet{gungor2024a,gungor2024b} showed that local PG disequilibration affects both the mean flow and turbulence of strong non-equilibrium APG TBLs, but with distinct impacts in the inner and outer regions.
Given that the inner region, primarily composed of small-scale turbulence, adjusts quickly to changes in the PG, it exhibits trends distinct from those in the outer region, which is dominated by large-scale turbulence \citep{gungor2024b}. 
Their study, however, could not isolate disequilibration effects from those of Reynolds number and accumulated PG history, which will be realized in this study for the first time.

Besides local disequilibration, developing an understanding of the upstream history effects is also crucial since they are present in most flows encountered in nature and industrial applications. 
A few studies have examined the impact of upstream PG history on flow statistics \citep{volinoNonequilibriumDevelopmentTurbulent2020,romeroPropertiesInertialSublayer2022,parthasarathy2023,preskett2025}.
For example,
\cite{marusicEvolutionZeropressuregradientBoundary2015} and \cite{schlatterTurbulentBoundaryLayers2012} studied history effects by examining the influence of different upstream BL tripping devices on the downstream development of a ZPG TBL.
On the other hand, \cite{bobkeHistoryEffectsEquilibrium2017} analyzed the impact of upstream PG history on the local boundary layer state by systematically comparing statistics from various simulation datasets. 
They were able to isolate the effects of upstream history through consideration of matched $\beta_{C}$ and $Re_\tau$ cases between different simulation data, thereby eliminating the effects of the local PG and Reynolds number. 
This enabled elucidation of the pressure-gradient history effects for APG TBLs at low $\beta_{C}$, which were later quantified by \cite{vinuesaRevisitingHistoryEffects2017} through proposal of the accumulated pressure-gradient parameter defined as:
\begin{equation}
    \overline{\beta}_{C,\theta} = \frac{1}{Re_{\theta} - Re_{\theta,0}}\int_{Re_{\theta,0}}^{Re_{\theta}} {\beta_{C}}\:(Re_{\theta})\;{{\textrm d}Re_{\theta}}
    \label{eq1}
\end{equation} 
where $Re_{\theta} = \theta U_e/\nu$ ($\theta$ is the momentum thickness and $U_e$ is the velocity at the edge of the boundary layer). 
\citet{vinuesaRevisitingHistoryEffects2017} found $\overline{\beta}_{C,\theta}$ to be effective for parameterizing the impact of accumulated pressure gradient \d{histories} on integral quantities, at least for their investigated regime of near-equilibrium TBLs at low-to-moderate $\beta_{C}$. 
The unique utility of this parameter is its ability to quantify the upstream flow history of an APG TBL through a single parameter. 
Additionally, $\overline{\beta}_{C,\theta}$ also enabled analytical modeling of APG skin friction coefficient ($C_f$) and shape factor ($H_{12}$) using ZPG statistics, making it a useful tool for comparing near-equilibrium boundary layers.
Recently, \cite{gomezLinearAnalysisCharacterizes2025} proposed a weighted mean of $\beta_{C}$ over streamwise distance $x$, as opposed to $Re_\theta$ used in (\ref{eq1}), to quantify accumulated pressure gradients. 
However, note here the inconsistency in definition of $\overline{\beta}_{C,\theta}$ (based on $Re_{\theta}$ or $x$) and matching flow conditions based on $Re_{\tau}$, which will also be addressed in this study.
Note that throughout this manuscript, $u$, $v$, and $w$ represent velocity fluctuations along the streamwise ($x$), wall-normal ($y$), and spanwise ($z$) directions respectively, while $U$, $V$, and $W$ indicate the respective mean velocities.

\begin{table*}[t]
    \centering
    \resizebox{\textwidth}{!}{
    \begin{tabular}{ccccccccc}
        \textbf{Database} & \textbf{TBL} & \textbf{$Re_{\tau}$} & \textbf{$Re_{\theta}$} & \textbf{$Re_{ZS}$} & \textbf{$\beta_{C}$} & \textbf{$\beta_{ZS}$} & \textbf{Colour shading} & \textbf{Reference} \\ \hline
        ZPG & LES ZPG & 150 - 2570 & 70 - 8300 & 100 - 1800 & - & - & \color{black}{black} & \cite{eitel-amorSimulationValidationSpatially2014a} \\ 
        b1.0 & LES APG & 150 - 863 & 80 - 3300 & 90 - 1200 & 8.3e-05 - 1.12 & -0.004 - 0.031 & \color{Goldenrod}{golden} & \cite{bobkeHistoryEffectsEquilibrium2017} \\ 
        b1.4 & LES APG & 150 - 2250 & 91 - 9650 & 100 - 3000 & 0.02 - 1.64 & 0.043 - 0.138 & \color{red}{red} & \cite{pozueloAdversepressuregradientTurbulentBoundary2022a} \\ 
        DNS16 & DNS APG & 10 - 450 & 900 - 4600 & 200 - 8400 & 1.7e-03 - 6.9e+04 & -0.013 - 0.467 & \color{darkgreen}{darkgreen} & \cite{gungorScalingStatisticsLargedefect2016b} \\ 
        DNS22 & DNS APG & 210 - 1000 & 1600 - 8700 & 400 - 12600 & 0.0016 - 422.88 & -0.053 - 0.307 & \color{blue}{blue} & \cite{gungorEnergyTransferMechanisms2022} \\ 
    \end{tabular}}
    \caption{Table summarizing the parametric space associated with the LES and DNS databases analyzed in this study. Definitions for $\beta_{C}$ and $\beta_{ZS}$ can be found in \ref{sec:methodology}. Here, $Re_{\tau} = U_{\tau}\delta/\nu$ and $Re_{ZS} = (\delta U_{ZS}/\nu)(U_{ZS}/U_e)$. The distribution of the pressure-gradient parameters over the Reynolds number can be found in Figure \ref{fig1}.}
    \label{tab1}
\end{table*}

While comparisons at matched $\beta_{C}$ over $Re_\tau$ have been shown to work well for weak-to-moderately strong APG TBLs, the quantification of PG strength by $\beta_C$ breaks down at strong APG conditions because the wall friction force becomes negligible.
Various theoretical approaches have thus been proposed to define alternatively scaled PG parameters based on specific regions of the TBL.
For instance,
\cite{castilloSeparationCriterionTurbulent2004} and \cite{castilloSimilarityAnalysisNonequilibrium2004} demonstrated that the \citet{zagarolaScalingMeanVelocity1997} scaling, ${U_{ZS}} = U_e \delta^*/\delta$, effectively regroups velocity defect profiles across a wide range of strong APG TBL cases.
\cite{maciel2006self} also identified Zagarola-Smits (ZS) scaling to be optimal for outer velocity scaling in strong APG TBLs, due to its ability to regroup defect profiles and Reynolds stresses \citep{maciel2006self} in strong APG TBLs, where conventional ZPG TBL scales like  $U_{\tau}$ and $U_e$ fail.
Building on this, \cite{macielOuterScalesParameters2018} laid the foundation for using the Zagarola-Smits scaled pressure-gradient parameter, $\beta_{ZS} = (\delta/(\rho U_{ZS}^2))(\partial{P}/\partial{x})_e$, explaining how force ratios can be expressed to construct PG strength parameter associated with the outer region.
This is consistent with the previous idea of \cite{glEquilibriumTurbulentBoundary}, who proposed the parameter $\beta_{i} = (\nu/(\rho U_{\tau}^3))(\partial{P}/\partial{x})_w$ to represent the inner-layer force balance, where $(\partial{P}/\partial{x})_w$ is the PG at the wall.
Similarly, \cite{maAsymptoticExpansionsScaling2024} documented other scales based on pressure-gradient strength \citep{weiMeanWallnormalVelocity2023,gungorScalingStatisticsLargedefect2016b,zagarolaScalingMeanVelocity1997, glEquilibriumTurbulentBoundary,pantonReviewWallTurbulence2005} and introduced an outer-scaled velocity that normalizes the outer peaks of APG TBL Reynolds stresses to define a PG parameter similar to $\beta_{ZS}$.   
More recently, \cite{gungor2024b} utilized $\beta_{ZS}$, $\beta_i$ as well as their local streamwise rates of change to unravel the distinctive reactions of the inner and outer regions to strong pressure-gradient local disequilibration effects and cumulative disequilibration effects.
Comparisons at matched $\beta_i$ and $\beta_{ZS}$ were also presented to understand the relative contributions of local and accumulated pressure gradients for strongly non-equilibrium APG TBLs, but without matching the Reynolds number.
Notably, the analysis of \cite{gungor2024b} found that the effects of local disequilibration (on the flow statistics) are distinct for momentum-loss and momentum-gain scenarios (\emph{i.e.}, corresponding to increasing or decreasing $\beta_{ZS}$ or $\beta_i$, respectively).  

To summarize this section, despite extensive research on the effects of APG in the past (as briefly introduced above), no study has isolated the impact of upstream PG history as well as local disequilibration across weak-to-strong APG strengths.
For instance, although \cite{bobkeHistoryEffectsEquilibrium2017} isolated the effects of history by focusing on matched $\beta_C$ and $Re_{\tau}$ cases, the analysis was limited to low-$\beta_C$ TBLs. 
On the other hand, \cite{gungor2024b} investigated the effects of history on non-equilibrium APG TBLs with high $\beta$, but without isolating the Reynolds number and local disequilibration effects. 
What is thus lacking is a comprehensive analysis that quantifies the effects of upstream history and local disequilibration over a wide range of $\beta$ parameters based on a choice of consistent scaling parameters.
This will be attempted here by drawing inspiration from the accumulated $\beta$-averaging method of \citet{vinuesaRevisitingHistoryEffects2017}, which will be imposed on the alternative PG parameters proposed by studies discussed above.
This forms the primary focus of the present study, which aims to propose a structured methodology to understand and quantify the effects of upstream PG history and local disequilibration. 
Inspired by the work of \cite{vinuesaRevisitingHistoryEffects2017} and \cite{gungor2024b}, we test using $\overline{\beta}_{ZS}$, defined as an average of $\beta_{ZS}$ over $Re_{ZS}$ or $Re_{\theta}$, to quantify accumulated outer-region pressure gradients for strong APG TBLs at matched $\beta_{ZS}$ and $Re_{ZS}$ or $Re_{\theta}$. 
Similarly, we test $\overline{\beta}_{C,\theta}$, defined as an average of $\beta_{C}$ over $Re_{\theta}$, for weak APG TBLs at matched $\beta_{C}$ and $Re_{\theta}$. 
It must be noted that these accumulated PG parameters, by construction, do not fully capture the flow history as they do not account for delayed turbulence response or the attenuation of pressure gradient effects with streamwise distance. 
Consequently, particular attention will be given to assessing the significance of this limitation.
Furthermore, the primary reason behind using $\beta_{ZS}$ instead of $\beta_C$ lies in the fact that $\beta_C$ does not capture the force balance for strong APG cases, whereas $\beta_{ZS}$ remains valid across all APG cases. \\
In the present work, we employ five published, high-fidelity ZPG and APG datasets \citep{bobkeHistoryEffectsEquilibrium2017, pozueloAdversepressuregradientTurbulentBoundary2022a, gungorScalingStatisticsLargedefect2016b, gungorEnergyTransferMechanisms2022} covering a wide range of $\beta$ and Reynolds numbers. 
Two of these four datasets correspond to near-equilibrium TBLs with weak to moderate APGs, while the other two are APG TBLs in strong non-equilibrium. 
Velocity variances and mean velocity contributions will be analyzed to assess the applicability of the history and local disequilibration parameters across broad APG strengths and equilibrium states.
In addition to offering several cases with matched $\beta$ and $Re$ (for appropriate choice of scaling), the datasets include a unique case where $\beta$ and $Re$ are matched across three TBLs, allowing us to closely study the effects of local disequilibration and accumulated PG.
Once established, the varying upstream history and disequilibration cases are used to understand effects on small- and large-scale velocity variance, achieved via scale-based energy decomposition.
Note that the symbol $\beta$ is used as a generic or umbrella notation for local pressure-gradient (PG) measures, with $\beta_C$ and $\beta_{ZS}$ representing specific scaled forms under this general definition.


\begin{figure*}[t]
     \centering
        \includegraphics[width=\textwidth]{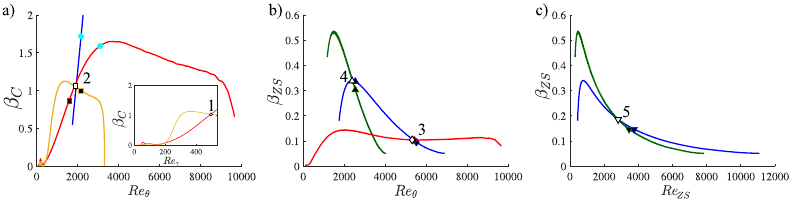}
        \caption{$\beta$ vs $Re$ plots illustrating the Exactly-Matched and Quasi-Matched cases across the four data sets. 
        Exactly-Matched cases for (a) $\beta_{C}$ vs $Re_{\theta}$ ($\beta_{C}$ vs $Re_{\tau}$ in the inset), (b) $\beta_{ZS}$ vs $Re_{\theta}$ and (c) $\beta_{ZS}$ vs $Re_{ZS}$ are shown using symbols: $\circ$- Case 1; $\square$- Case 2; $\diamond$- Case 3; $\bigtriangleup$- Case 4 and $\bigtriangledown$- Case 5. Filled {\color{cyan}cyan} circles indicate Case DISEQ.
        Line plots correspond to datasets mentioned in Table \ref{tab1}. 
        The white-filled symbols indicate the Exactly-Matched cases while the symbols filled in black represent Quasi-matched cases, which were selected due to the unavailability of spectral data at the location of the white-filled symbols.
        Further details about these cases can be found in Table \ref{tab2} and in $\S$\ref{subsec:analysis methodology}.}
         
        \label{fig1}
\end{figure*}

\begin{table*}[h!]
    \centering
    \renewcommand{\arraystretch}{0.8}
    \setlength{\tabcolsep}{4pt}

    \begin{tabular*}{\linewidth}{@{\extracolsep{\fill}} l l l l ccccc ccccc}
        \toprule
        \textbf{Case\#} & \textbf{Dataset} & \textbf{Line Style} & \multicolumn{5}{c}{\textbf{Exactly-Matched $\beta$, $Re$}} & \multicolumn{5}{c}{\textbf{Quasi-Matched $\beta$, $Re$}} \\
        \cmidrule(lr){4-8} \cmidrule(lr){9-13}
        \textbf{} & \textbf{} & \textbf{} &
        $\mathbf{Re_{\tau}}$ & $\mathbf{\beta}_{C}$ & $\mathbf{\overline{\beta}_{C,\theta}}$ & $\mathbf{d\beta_C/dRe_{\tau}}$ & $\mathbf{H_{12}}$ &
        $\mathbf{Re_{\tau}}$ & $\mathbf{\beta_C}$ & $\mathbf{\overline{\beta}_{C,\theta}}$ & $\mathbf{d\beta_C/dRe_{\tau}}$ & $\mathbf{H_{12}}$ \\
        \midrule

        \multirow{2}{*}{\textbf{Case 1}} 
        & b1.0   & \color{Goldenrod}{\sampleline[solid]{1pt}} & 494  & 1.02  & 0.83  & -1.1e-03 & 1.60  & -  & - & - & - & - \\  
        & b1.4   & \color{red}{\sampleline[solid]{1pt}}      & 494  & 1.02  & 0.48  & 3.5e-03  & 1.56  & - & - & - & - & - \\  
        \midrule

        \textbf{} & \textbf{} & \textbf{} &
        $\mathbf{Re_{\theta}}$ & $\mathbf{\beta_C}$ & $\mathbf{\overline{\beta}_{C,\theta}}$ & $\mathbf{d\beta_C/dRe_{\theta}}$ & $\mathbf{H_{12}}$ &
        $\mathbf{Re_{\theta}}$ & $\mathbf{\beta_C}$ & $\mathbf{\overline{\beta}_{C,\theta}}$ & $\mathbf{d\beta_C/dRe_{\theta}}$ & $\mathbf{H_{12}}$ \\
        \midrule

        \multirow{3}{*}{\textbf{Case 2}} 
        & DNS22 & \color{blue}{\sampleline[solid]{1pt}}     & 1927 & 1.04 & 0.55 & 2.8e-03 & 1.46 & - & -  & -  & - & - \\  
        & b1.4  & \color{red}{\sampleline[solid]{1pt}}      & 1860 & 1.06 &  0.50 &  6.4e-04 &  1.56 & 1580 & 0.86  & 0.39  & 7.4e-04 & 1.55 \\  
        & b1.0  & \color{Goldenrod}{\sampleline[solid]{1pt}}& 1920 &  1.04 &  0.82 &  -1.8e-04 &  1.61 & 2152 & 1.00  & 0.84  & -3.6e-04 & 1.60 \\  
        \midrule

        \textbf{} & \textbf{} & \textbf{} &
        $\mathbf{Re_{\theta}}$ & $\mathbf{\beta_{ZS}}$ & $\mathbf{\overline{\beta}_{ZS,\theta}}$ & $\mathbf{d\beta_{ZS}/dRe_{\theta}}$ & $\mathbf{H_{12}}$ &
        $\mathbf{Re_{\theta}}$ & $\mathbf{\beta_{ZS}}$ & $\mathbf{\overline{\beta}_{ZS,\theta}}$ & $\mathbf{d\beta_{ZS}/dRe_{\theta}}$ & $\mathbf{H_{12}}$ \\
        \midrule

        \multirow{2}{*}{\textbf{Case 3}} 
        & DNS22 & \textcolor{blue}{\sampleline[dotted]{1pt}}& 5373 & 0.103  & 0.20  & -6e-05 & 2.42 & 5521.87 & 0.10  & 0.22  & -6.1e-05 & 2.50 \\  
        & b1.4  & \color{red}{\sampleline[dotted]{1pt}}     & 5373 & 0.106  & 0.12  & -4.3e-06 & 1.60 & 5418.14 & 0.10  & 0.11  & -3.2e-06 & 1.60 \\  
        \midrule

        \multirow{2}{*}{\textbf{Case 4}} 
        & DNS22 & \color{blue}{\sampleline[dashed]{1pt}}    & 2404 & 0.34  & 0.28  & -1.4e-05 & 1.54 & 2541.59 & 0.33  & 0.26  & -2.8e-05 & 1.56 \\  
        & DNS16 & \color{darkgreen}{\sampleline[dashed]{1pt}} & 2404 & 0.34  & 0.44  & -8.4e-04 & 1.92 & 2540.79 & 0.30  & 0.40  & -8.3e-04 & 2.00 \\  
        \midrule

        \textbf{} & \textbf{} & \textbf{} &
        $\mathbf{Re_{ZS}}$ & $\mathbf{\beta_{ZS}}$ & $\mathbf{\overline{\beta}_{ZS,ZS}}$ & $\mathbf{d\beta_{ZS}/dRe_{ZS}}$ & $\mathbf{H_{12}}$ &
        $\mathbf{Re_{ZS}}$ & $\mathbf{\beta_{ZS}}$ & $\mathbf{\overline{\beta}_{ZS,ZS}}$ & $\mathbf{d\beta_{ZS}/dRe_{ZS}}$ & $\mathbf{H_{12}}$ \\
        \midrule

        \multirow{2}{*}{\textbf{Case 5}} 
        & DNS22 & \color{blue}{\sampleline[dash pattern=on .7em off .2em on .05em off .2em]{1pt}} & 3250 & 0.16  & 0.25  & -3.2e-05 & 2.10 & 3760.4 & 0.15  & 0.23  & -2.3e-05 & 2.15 \\  
        & DNS16 & \color{darkgreen}{\sampleline[dash pattern=on .7em off .2em on .05em off .2em]{1pt}} & 3250 & 0.16  & 0.31  & -7e-05 & 2.50 & 3467.1 & 0.15  & 0.30  & -6.6e-05 & 2.56 \\  
        \midrule
        \multicolumn{13}{c}{\textbf{Case DISEQ : Not Matched (included for comparative analysis)}} \\
        \midrule
        \textbf{} & \textbf{Dataset} & \textbf{Line Style} &
        $\mathbf{Re_{\theta}}$ & $\mathbf{\beta_{C}}$ & $\mathbf{\overline{\beta}_{C,\theta}}$ & $\mathbf{d\beta_C/dRe_{\theta}}$ & $\mathbf{H_{12}}$ \\
        \midrule
        \multirow{2}{*}{\textbf{Case DISEQ}} & DNS22 & \color{blue}{\sampleline[solid]{1pt}} & 2170 & 1.72 & 0.89 & 2.9e-03 & 1.50 \\
        \textbf{}  & b1.4  & \color{red}{\sampleline[solid]{1pt}}  & 3098 & 1.59 & 0.89 & 2.2e-04 & 1.61 \\
        \bottomrule

    \end{tabular*}
    \caption{Summary of the various cases analyzed in this study, with Case\# corresponding to $\beta$ and $Re$ marked in Figure~\ref{fig1}. The "Exactly-Matched" subgroup indicates the parametric values for comparison of mean velocities and variances in subsequent analysis, while "Quasi-Matched" subgroup contains the parametric values considered for comparing the scale-decomposed variances, due to the unavailability of spectral energy data at "Exactly-matched" cases. Only relevant parameters for each case are included. Different line styles have been considered for Cases 3 to 5 to enhance comprehension in Figures~\ref{fig5} and~\ref{fig6}. Note that Case DISEQ is not a matched case and is included to study the effects of local disequilibration on turbulent scales.}
    \label{tab2}
\end{table*}

\section{Methodology}
\label{sec:methodology}
\subsection{Databases} %

We employ five published high-fidelity TBL databases---three computed using large-eddy simulations (LES) and two based on direct numerical simulations (DNS).
Note that all three LES databases are highly resolved, with a resolution only two times coarser than a conventional DNS in the streamwise and spanwise directions, thus ruling out any influence on the conclusions of the present study. 
The unresolved turbulent kinetic energy dissipation for the LES (which is around 10\% of the total) is accounted for by adding a small body force, resulting in minimal differences compared to a conventional DNS simulation (refer to \citealp{eitel-amorSimulationValidationSpatially2014a} for further details).
Here, we consider an LES database from \cite{eitel-amorSimulationValidationSpatially2014a} for a ZPG TBL spanning across all $Re_{\tau}$ values considered in the other four APG TBL databases. 
Among these APG datasets, two correspond to near-equilibrium APG TBLs with weak to moderate pressure gradients \citep{bobkeHistoryEffectsEquilibrium2017,pozueloAdversepressuregradientTurbulentBoundary2022a}, while the other two represent APG TBLs in strong disequilibrium, featuring regions of strong APG \citep{gungorScalingStatisticsLargedefect2016b,gungorEnergyTransferMechanisms2022}. 
These datasets enable the systematic study of the effects of upstream PG history and local disequilibration.
The selected databases cover a wide range of Reynolds numbers and $\beta$ parameters, with their relevant parametric space summarized in table \ref{tab1}. 
The two DNS databases, referred to as DNS16 and DNS22, were originally set up to impose a rapidly increasing adverse PG followed by a decreasing APG.
While the two LES databases, b1.0 and b1.4, are set up to establish a near-equilibrium state at $\beta_{C}$ values of 1 and 1.4, respectively, after a gradual increase in PG strength with downstream development of the TBL. 
Here, a state of near-equilibrium is considered achieved when $\beta_{C}$ remains approximately constant over a considerable streamwise distance.

The mean boundary layer thicknesses ($\delta$) for all TBL profiles have been computed using the diagnostic plot method, as documented in \cite{vinuesaDeterminingCharacteristicLength2016b}.
While we recognize the existence of alternate methods to estimate $\delta$ \citep{lozier2025}, its exact definition does not influence the present conclusions.

\subsection{Case Selection and Analysis Methodology}
\label{subsec:analysis methodology}

Here, we draw inspiration from \citet{bobkeHistoryEffectsEquilibrium2017} and use the four APG TBL databases to identify matched $\beta$ and Re cases to limit the differences between them solely to upstream PG history and local disequilibrating effects.
The selected cases (illustrated in Figure \ref{fig1}) are based on three differently scaled Reynolds numbers ($Re_{\tau}$, $Re_{\theta}$, and $Re_{ZS}$) and two pressure-gradient parameters ($\beta_C$ and $\beta_{ZS}$) to assess their ability in truly isolating the aforementioned effects \citep{bobkeHistoryEffectsEquilibrium2017,maciel2006self,gungor2024b,gungor2024a}.
Table \ref{tab2} documents the values of relevant parameters for the "Exactly-Matched" and "Quasi-Matched" cases considered for the present analysis, with the differences between them obvious from the nomenclature.
While the former are considered for comparing mean velocities and variances in the subsequent analysis, the latter subgroup is considered for comparing the energy spectra and scale-decomposed variances due to the unavailability of spectral energy data for the "Exactly-matched" cases
(which are typically only saved at certain streamwise locations of the domain).

Case 1 is a well-established comparison between the two low-$\beta$ APG TBLs b1.4 and b1.0 \citep{bobkeHistoryEffectsEquilibrium2017,pozueloAdversepressuregradientTurbulentBoundary2022a} at matched $\beta_{C}$ and $Re_{\tau}$ while ${\rm d}\beta_C/{\rm d}Re_{\tau}$ and $\overline{\beta}_{C,\theta}$ differ (refer to the inset in Figure \ref{fig1}). 
Case 1 essentially revisits the effects of history on mean velocity, $\overline{u^2}$ and $\overline{v^2}$ variances evaluated in the literature by \cite{bobkeHistoryEffectsEquilibrium2017} and \cite{pozueloAdversepressuregradientTurbulentBoundary2022a}. 
However, the definition of $\overline{\beta}_{C,\theta}$ (based on $Re_{\theta}$) and matched parameter (based on $Re_{\tau}$) have an inconsistent choice of Reynolds numbers. This is especially relevant for high-$\beta$ cases where the $Re_{\tau}$ definition falls apart. Due to this, we refrain from making inferences on scale-decomposed variances based on Case 1. 
We will therefore consider Case 2 at matched $\beta_C$ and $Re_{\theta}$ to evaluate history effects based on $\overline{\beta}_{C,\theta}$.

Case 2 examines a distinctive comparison where two low-$\beta_{C}$ boundary layers (b1.4 and b1.0) and a highly non-equilibrium APG boundary layer (DNS22) have matched values of $\beta_{C}$ and $Re_{\theta}$. 
This case offers a unique opportunity to reasonably isolate the effects of local disequilibration, emerging from significant differences in $d\beta_C/dRe_{\tau}$ between the b1.4 and DNS22 TBLs, while maintaining only minor differences in $\overline{\beta}_{C,\theta}$. Comparisons of b1.0 with DNS22 and b1.4 bolster our observations of the combined effects of local disequilibration and accumulated PG.

Cases 3 and 4 are selected based on matching $\beta_{ZS}$ and $Re_{\theta}$. 
The PG parameter $\beta_{C}$ has been replaced by $\beta_{ZS}$ because, as noted in the Introduction, $\beta_{C}$ fails to account for the pressure force contribution in the force balance for large-velocity-defect APG TBLs (typically when $H_{12} > 1.8$). In contrast, $\beta_{ZS}$ remains valid across all APG conditions. The Reynolds number $Re_{\theta}$ is used due to its widespread adoption as the local Reynolds number in boundary layer research. However, in Case 5, we adopt a further consistent set of parameters --- $\beta_{ZS}$ and $Re_{ZS}$.

Case 3 compares relatively downstream locations of the numerical domains associated with b1.4 and DNS22, while Case 4 compares the two high-$\beta$ non-equilibrium datasets in their relatively upstream regions, where the local pressure force impact, quantified by $\beta_{ZS}$, is high. 
Case 5 allows us to compare two non-equilibrium boundary layers at matched $\beta_{ZS}$ and $Re_{ZS}$ values, hence utilizing the Zagarola-Smits scaling for both pressure-gradient and Reynolds number.  For reference, the $Re_{\theta}$ values corresponding to the Exactly-Matched column in Case 5 are 4481 for DNS22 and 3175 for DNS16, while the Quasi-Matched column corresponds to 4737 and 3229, respectively.

Case DISEQ represents a special scenario with fairly matched $\beta_C$ values, identical $\overline{\beta}_{C,\theta}$ values, and local disequilibration values that differ by an order of magnitude. While the $Re_{\theta}$ values differ (2200 for DNS22 and 3100 for b1.4), trends in the small-scale near-wall region can be attributed to local disequilibration due to the region's sensitivity to PG changes. Case DISEQ represents a location that is not matched in $\beta$ and $Re$, but enables a more focused examination of local disequilibration effects on turbulent scales. An in-depth discussion on the power-spectral density in Case DISEQ is provided in Section \ref{subsec:low defect TBLs}.

Note that $Re_{ZS}$ and $\beta_{ZS}$ are obtained based on a momentum equation analysis demonstrated previously in \cite{macielOuterScalesParameters2018}, where the order of magnitude of the Reynolds shear stress gradient term in the outer region is kept constant in the momentum equation. The choice of the $U_{ZS}$ velocity scale results in $Re_{ZS}$ and $\beta_{ZS}$ as non-dimensional parameters that accurately follow the force ratios in the outer region.

It must be noted that although here we are using $Re_{\theta}$ as a suitable parameter for TBLs with large-defects, it does not have a rigorous theoretical formulation, unlike $Re_{ZS}$.
The similarities and differences observed (if any) between matched cases using these two different Reynolds number definitions will be discussed in $\S$\ref{sec:results}.
Note that for Cases 3 and 4, where the cases are matched based on $Re_{\theta}$, the definition for the accumulated $\beta_{ZS}$ follows equation (\ref{eq2}), while Case 5 being a matched case of $Re_{ZS}$ is defined by equation (\ref{eq3}) following:
\begin{equation}
    \overline{\beta}_{ZS,\theta} = \frac{1}{Re_{\theta} - Re_{\theta,0}}\int_{Re_{\theta,0}}^{Re_{\theta}}\beta_{ZS}(Re_{\theta})\:{\textrm d}Re_{\theta}, \textrm{and} 
    \label{eq2}
\end{equation} 
\begin{equation}
    \overline{\beta}_{ZS,ZS} = \frac{1}{Re_{ZS} - Re_{ZS,0}}\int_{Re_{ZS,0}}^{Re_{ZS}}\beta_{ZS}(Re_{ZS})\:{\textrm d}Re_{ZS}.
    \label{eq3}
\end{equation}
Averaging in this way aims at capturing the nature of upstream development of the APG TBLs over the relevant Reynolds number range, thus quantifying their accumulated history. For the b1.0 and b1.4 datasets, the lower limit $Re_{\theta,0}$ is chosen at a slightly downstream location from the start of their simulation domains ($Re_{\theta}$ $\sim$ $320$), to ensure that the boundary layer is fully turbulent and under ZPG condition. 
On the other hand, this lower limit is at the start of the DNS22 and DNS16 simulation domains, where the boundary layer is already fully turbulent.
\begin{figure*}[t!]
    \centering
    \includegraphics[width= \textwidth]{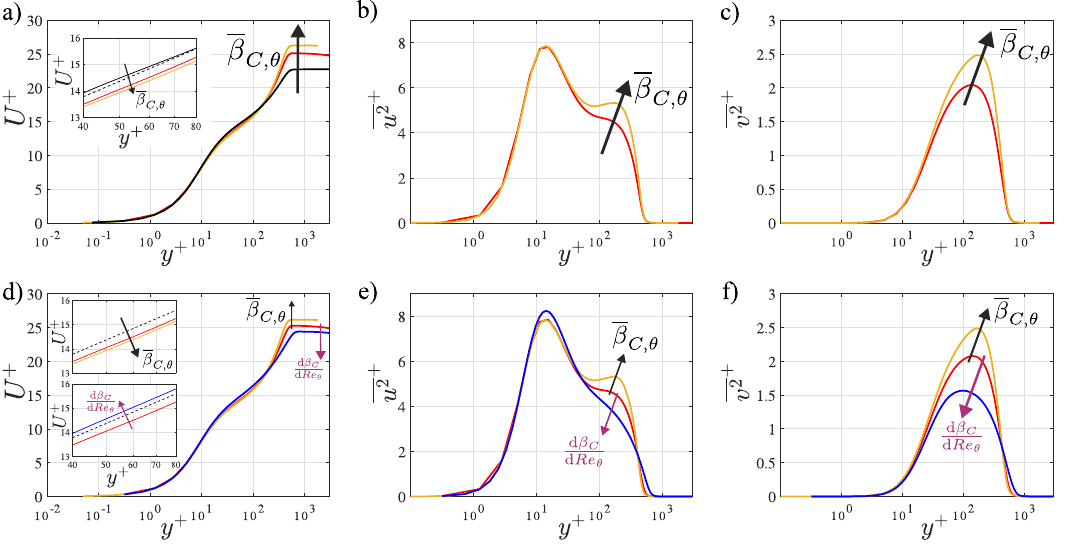}
    \caption{Mean Statistics and Velocity variances for Cases 1 (top) and 2 (bottom). All quantities are scaled in viscous units - $U_{\tau}$ and $l_{\tau}$. (a,d) Streamwise mean velocity, (b,e) $\overline{u^2}$ variances, (c,f) $\overline{v^2}$ variances. 
    Colors and Line styles correspond to those in Tables \ref{tab1} and \ref{tab2}.
    Solid black lines in (a,d) correspond to the ZPG TBL profiles at matched $Re_{\tau}$ and $Re_{\theta}$ respectively, while the dashed line in (a) represents the logarithmic law.}
    \label{fig2}
\end{figure*}

Availability of spectral energy data at the "Quasi-Matched" cases plays a crucial role in this study, enabling analysis of small- and large-scale energy contributions as a function of upstream PG history.
Although the $\beta$ and $Re$ values do not match exactly for these comparisons, the slight mismatch does not influence the statistical trends inferred in subsequent sections.
We utilize the cutoff based on the viscous-scaled spanwise wavelength ($\lambda^+_{z,c} = 300$) proposed by \cite{deshpandeReynoldsnumberEffectsOuter2023d} to decompose the total spectral energy at each wall-normal location into small ($\overline{u^2_{SS}}$; $\lambda^+_{z,c}$ $\le$ 300) and large scale contributions ($\overline{u^2_{LS}}$; $\lambda^+_{z,c}$ $>$ 300), such that $\overline{u^2}$($y^+$) = $\overline{u^2_{SS}}$($y^+$) + $\overline{u^2_{LS}}$($y^+$), and $\overline{v^2}$($y^+$) = $\overline{v^2_{SS}}$($y^+$) + $\overline{v^2_{LS}}$($y^+$). 
\cite{deshpandeReynoldsnumberEffectsOuter2023d} arrived at this cutoff value iteratively, based on the well-accepted argument of viscous-scaled small-scale energy ($\overline{u^2_{SS}}^+$) exhibiting $Re_{\tau}$-invariance for $\lambda_{z,c}^+ \le 300$ (for the case of canonical flows; refer to figure 12 of \citealp{deshpandeReynoldsnumberEffectsOuter2023d}).
Since this spanwise wavelength cutoff, $\lambda^+_{z,c}$ is constructed based on a viscous-scaled parameter that breaks down for strong APG TBLs, it becomes ineffective in higher-defect cases. 
This limitation is overcome by utilizing an outer-scaled spanwise wavelength ($\lambda_{z,c}/\delta$), the cutoff value for which is estimated iteratively such that the small- and large-scale velocity variance profiles nominally match those computed based on the established value of $\lambda^+_{z,c}$ = 300, for the low-defect Cases 1 and 2.
The cutoff value of $\lambda_{z,c}/\delta$ = 0.5 shows reasonable agreement between the scale-decomposed variances, as demonstrated in the plots provided in Appendix \hyperref[sec:appendix_A]{A}.


\section{Results}
\label{sec:results}

We now analyze the five matched cases and the special Case DISEQ described in the previous section to understand the upstream PG history and local disequilibration effects.
The primary goals are to: a) establish the effect of $d\beta_C/dRe_{\tau}$ on the flow statistics and b) to determine whether the accumulated PG parameters (defined in equations \ref{eq1}, \ref{eq2} and \ref{eq3}) capture the upstream PG history effects across a broad range of pressure gradient strengths.
The success of the latter will facilitate establishing the effects of PG history on the small- and large-scale energies across weak-to-strong APG TBLs.

Note here that for low-to-moderately strong APG cases (characterized by $\beta_{C}$),
the profiles of velocity statistics are scaled with the friction velocity scale $U_{\tau} = \sqrt{(\tau_w/\rho)}$ and length scale $l_{\tau} = \nu/U_{\tau}$ (where $\rho$ is the fluid density).
Conversely, for strong APG cases (quantified by $\beta_{C}$), outer scaling with velocity scale $U_e$ and length scale $\delta$ is utilized due to the breakdown of wall-scaling with increase in APG strength.

\subsection{Low-to-moderately strong APG TBLs} 
\label{subsec:low defect TBLs}
We begin with Case 1 comparing the TBL datasets b1.4 and b1.0, analyzed previously by \citet{pozueloAdversepressuregradientTurbulentBoundary2022a}, to revisit the pre-established ability of the $\overline{\beta}_{C,\theta}$ parameter to capture the effects of APG history for moderate values of $\beta_C$ \citep{vinuesaRevisitingHistoryEffects2017}.
Figure \ref{fig2}(top) depicts the inner-scaled mean velocity profiles along with the inner-scaled $\overline{u^2}$ and $\overline{v^2}$ variances for the two data sets at Exactly-Matched Case 1. 
As already discussed by \citet{pozueloAdversepressuregradientTurbulentBoundary2022a}, the behavior of the boundary layer statistics and the greater turbulence energisation observed in the b1.0 simulation are logically accounted for by $\overline{\beta}_{C,\theta}$, consistent with the original description in \citet{bobkeHistoryEffectsEquilibrium2017}.
The mean velocity profiles reveal two key observations that illustrate the effects of upstream history. 
First, b1.0 having $\overline{\beta}_{C,\theta}$ of 0.83, which is more than twice that of the b1.4 (0.48), is characterized by a greater deviation from the ZPG in the log region, as seen in the inset of Figure \ref{fig2}(a).
Similar deviations from the classical log law with increasing APG strength have been reported across low-to-high Reynolds-number TBLs in the past \citep{nagibVariationsKarmanCoefficient2008d, montyParametricStudyAdverse2011, indinger_mean-velocity_2006, knoppExperimentalAnalysisLog2021, pozueloAdversepressuregradientTurbulentBoundary2022a,
deshpandeReynoldsnumberEffectsOuter2023d,gungor2024a}. 
Second, in the outer region, the b1.0 TBL exhibits a larger mean velocity deficit as well as higher Reynolds normal stress levels (Figures \ref{fig2}b and \ref{fig2}c), further highlighting the stronger cumulative effect of the APG associated with stronger history. 
Here, for brevity, we have limited our analysis only to $\overline{u^2}$ and $\overline{v^2}$ but similar trends are noted for $\overline{w^2}$ and $\overline{uv}$. 
Local disequilibration appears to play a minor role in Case 1 (relative to accumulated PG history), as the negative ${\rm d}\beta_C/{\rm d}Re_{\tau}$ observed for flow b1.0 would typically indicate a reduction in the APG effect \citep{gungor2024b}.
\begin{figure*}[t!]
    \centering
    \includegraphics[width = \textwidth]{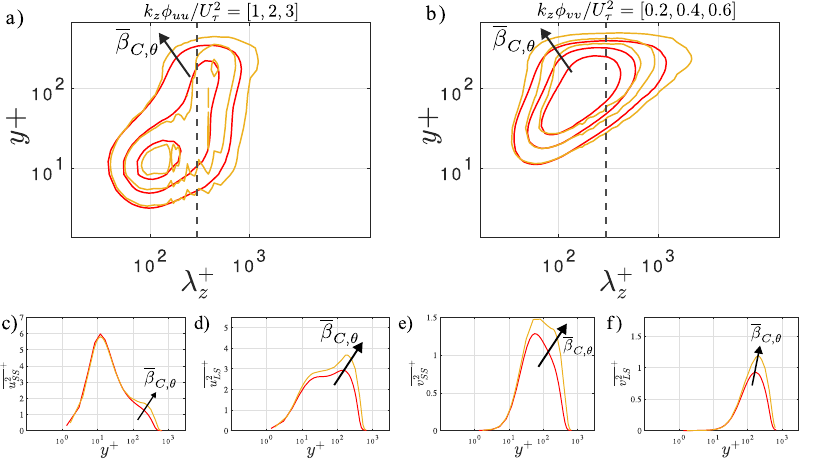}

    \caption{Premultiplied spanwise power-spectral density of the (a) streamwise, (b) wall normal velocity fluctuations ($k_z\phi_{uu}^+$ and $k_z\phi_{vv}^+$ respectively), small and large-scale contributions to velocity variances in Case 2 scaled in inner units. DNS22 contributions are not plotted due to the unavailability of power-spectral density data at a reasonable "Quasi-matched" location. Decomposition into small and large scales is done based on the $\lambda_{z,c}^+ = 300$ cutoff documented in \cite{deshpandeReynoldsnumberEffectsOuter2023d}.  All quantities are scaled in viscous units - $U_{\tau}$ and $l_{\tau}$. Figure also presents (c) Small scale contributions to $\overline{u^2}$ variances, (d) Large scale contributions to $\overline{u^2}$ variances, (e) Small scale contributions to $\overline{v^2}$ variances, (f) Large scale contributions to $\overline{v^2}$ variances. Colors and Line styles correspond to those in Tables \ref{tab1} and \ref{tab2}.}
    \label{fig3}
\end{figure*}
While Case 1 limits itself to a comparison of two low-$\beta$ TBLs at a matched $\beta_{C}$ and $Re_{\tau}$, in Case 2 we compare three TBLs at matched $\beta_C$ and $Re_{\theta}$ -- one non-equilibrium TBL DNS22, in its region of rapidly increasing PG and two low-$\beta$ TBLs b1.4 and b1.0.
The comparison in Figure 2 (bottom row) between the b1.0 TBL ($\overline{\beta}_{C,\theta} = 0.82$) and the b1.4 TBL ($\overline{\beta}_{C,\theta} = 0.50$) is consistent with the conclusion drawn from Case 1 regarding the influence of accumulated pressure gradient on TBL statistics, even when matching $Re_{\theta}$ instead of $Re_{\tau}$. We observe a greater mean velocity deficit, greater deviation from ZPG in the log region, and higher Reynolds normal stress levels in the outer region of the b1.0 TBL compared to the b1.4 TBL.
Figure \ref{fig3} presents premultiplied spanwise power-spectral density and scale-decomposed fluctuations scaled in inner units for Quasi-matched values associated with Case 2 (Table \ref{tab2}) to analyze the response of the turbulent scales to the changing PG conditions. The small- and large-scale contributions to $\overline{u^2}$ and $\overline{v^2}$ variances are decomposed based on the $\lambda^+_{z,c} = 300$ cutoff. Here, we only plot the power-spectral density and scale-decomposed variances for b1.0 and b1.4 due to the unavailability of a reasonable "Quasi-matched" location for DNS22. Note that the slight mismatch in $\beta$ and $Re$ values for Quasi-matched cases does not influence the statistical trends, and can be confirmed by referring to Appendix \hyperref[sec:appendix_B]{B}. Figure \ref{fig3}(a) shows the development of an outer peak in the $\overline{u^2}$ spectra of the b1.0 TBL, which is consistent with its greater $\overline{\beta}_{C,\theta}$ value, indicating a more pronounced APG effect. Furthermore, the energy in the outer region of the b1.0 spectra (both $\overline{u^2}$ and $\overline{v^2}$ energy) extends to both smaller and larger scales, further supporting this.

Now, observing $\overline{u^2_{SS}}$ and $\overline{u^2_{LS}}$ reveals a striking difference between small- and large-scale energisation in response to increasing accumulated PG ($\overline{\beta}_{C,\theta}$).
Small-scale contributions to $\overline{u^2}$ variances (see Figure \ref{fig3}a) exhibit a collapse in the inner region but a clear energisation in the outer region, aligning with the increasing $\overline{\beta}_{C,\theta}$ value. 
On the other hand, the large-scale contributions ($\overline{u^2_{LS}}$) plotted in Figure \ref{fig3}(b) unveil an energisation of large-scale variance contributions with increasing APG history across the entire TBL. Scale-decomposed contributions of $\overline{v^2}$ variances also demonstrate an energisation with $\overline{\beta}_{C,\theta}$ beyond the near-wall region, where most of the $\overline{v^2}$ energy is concentrated (Figures \ref{fig3}c,d).
While $\overline{u^2}$ is strong in both the near-wall and outer regions, the majority of $\overline{v^2}$ energy is concentrated in the outer region of the TBL. This is attributed to the dominance of large-scale energy transfer through the pressure-strain mechanism in the outer layer, as demonstrated by \cite{gungorEnergyTransferMechanisms2022} and \cite{deshpandeStreamwiseEnergytransferMechanisms2024}. Inner-scaled turbulence statistics in the lower-defect cases (Cases 1 and 2) show that accumulated PG and local disequilibration primarily affect the outer region, while the near-wall statistics remain largely unchanged. Since the $\overline{v^2}$ footprint is confined mainly to the outer layer, the entire wall-normal extent of $\overline{v^2}$ is sensitive to the effects of accumulated PG and local disequilibration.
\begin{figure*}[t]
     \centering
        \includegraphics[width=\textwidth]{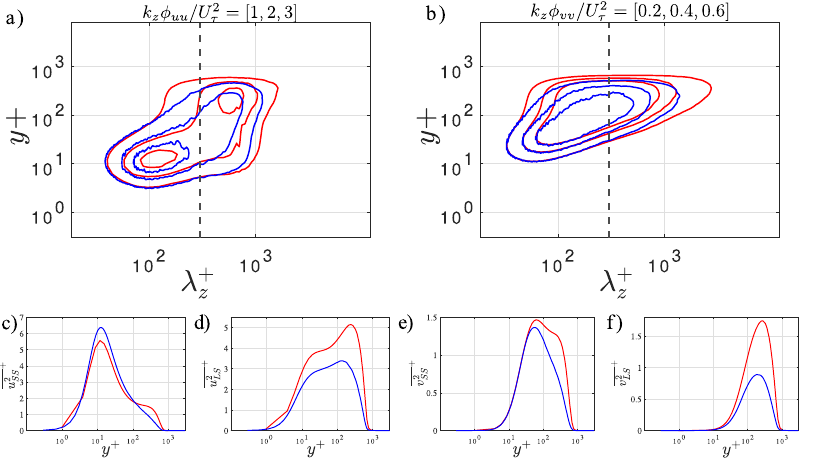}
        \caption{Contours of the premultiplied spanwise power-spectral density of (a) streamwise Reynolds stress scaled with friction velocity $k_z\phi_{uu}/U_{\tau}^2$, and (b) wall normal Reynolds stress scaled with friction velocity $k_z\phi_{vv}/U_{\tau}^2$ corresponding to Case DISEQ as per Table \ref{tab2}. Figure also presents (c) Small scale contributions to $\overline{u^2}$ variances, (d) Large scale contributions to $\overline{u^2}$ variances, (e) Small scale contributions to $\overline{v^2}$ variances, (f) Large scale contributions to $\overline{v^2}$ variances. The dotted line corresponds to the $\lambda_{z,c}^+ = 300$ scale cutoff as per \cite{deshpandeReynoldsnumberEffectsOuter2023d}. Colors and Line styles correspond to those in Tables 1 and 2.}
         
        \label{fig4}
\end{figure*} 

Uniquely Case 2 also allows us to study the local-\newline disequilibriating effects through the comparison between the b1.4 TBL and DNS22 TBL at a matched $Re_{\theta}$ and $\beta_{C}$.
With $\overline{\beta}_{C,\theta}$ being larger for DNS22, APG effects should have felt stronger for DNS22, which is opposite to what results show.
It is evident that the mean flow and turbulence response to the increasing APG, for DNS22, is significantly delayed compared to that for b1.4.
This delayed response of DNS22 is caused by the rapid increase in APG strength (\emph{i.e.} high ${\rm d}{\beta}/{\rm d}{Re_{\theta}}$). The flow is yet to fully adjust to the local $\beta$ recorded.
Mean velocity plots in Figure \ref{fig2}(d) show a relatively lower velocity deficit in the wake for DNS22 and less deviation from the log law compared to the b1.4 case (see inset of Figure \ref{fig2}d).
$\overline{u^2}$ and $\overline{v^2}$ levels in Figure \ref{fig2}(e,f) are also lower in the outer region for DNS22 but marginally higher in the inner region compared to that of b1.4, which can be associated with the unique effects of local disequilibration. 
These differing trends in the near-wall and outer region energy are consistent with the observations of \cite{gungor2024b}, who explained them based on the fact that near-wall scales respond rapidly to local disequilibrium while the outer region exhibits a slower reaction.

Another interesting consequence of the delayed response is the limited streamwise extent available for the DNS22 boundary layer across which it can adjust/redistribute energy from $\overline{u^2}$ to transverse $\overline{v^2}$ (see Figure \ref{fig2}f), facilitated via the pressure strain terms \citep{gungorEnergyTransferMechanisms2022,deshpandeStreamwiseEnergytransferMechanisms2024}. 
The delay in this turbulence energy transfer results in lower $\overline{v^2}$ levels in the outer region which causes the outer peak to be closer to the wall for DNS22 compared to b1.4.

Case 2 thereby highlights the limitation of the accumulated PG parameter $\overline{\beta}$ to account for delayed turbulence response and the attenuation of PG effects with streamwise distance. DNS22 statistics experience weaker APG effects than b1.4 and b1.0 due to the rapid increase in PG and the nascent stages of development of the DNS22 TBL.

To further understand the effects of local disequilibration, particularly on the turbulent scales, we analyze the premultiplied spanwise power-spectral density scaled in inner units for Case DISEQ in Figure \ref{fig4} (Table \ref{tab2}). 

The impact of local disequilibration is most evident by observing the differential behavior of small- and large-scale energy contours (left and right of the dotted line marking $\lambda^+_{z,c} = 300$ respectively). As shown in Figure \ref{fig4}(a), the $\overline{u^2}$ spectra for DNS22 exhibits a more energized inner peak than b1.4, indicating the rapid response of near-wall small scales to local PG changes. 
Meanwhile, the outer region of both $\overline{u^2}$ and $\overline{v^2}$ spectra (which includes both small and large scales), and the large scales in the inner region, reflect the delayed turbulence response associated with high disequilibration. The scale-decomposed variances shown in Figure \ref{fig4} (bottom row) further support the spectral observations, with DNS22 $\overline{u^2_{SS}}^+$ variances (in Fig \ref{fig4}(c)) displaying near-wall energization but a delayed response in the outer region.
These contrasting responses of the near-wall and outer regions are due to differences in their dominant turbulence scales; the near-wall region is dominated by physically small scales which have small time scales that respond quickly to any change in the PG conditions \citep{bobkeHistoryEffectsEquilibrium2017}.
On the other hand, large scales that predominate in the outer region have a longer response time scale, leading to delayed adaptation.
An exception, however, is the behaviour of small-scales in the outer region, which do not react as quickly to disequilibration as those in the near-wall region. 
Understanding the physical reasoning behind this warrants further investigation into their behavior, which is, however, beyond the present scope.

\begin{figure*}[t!]
    \centering
    \includegraphics[width= \textwidth]{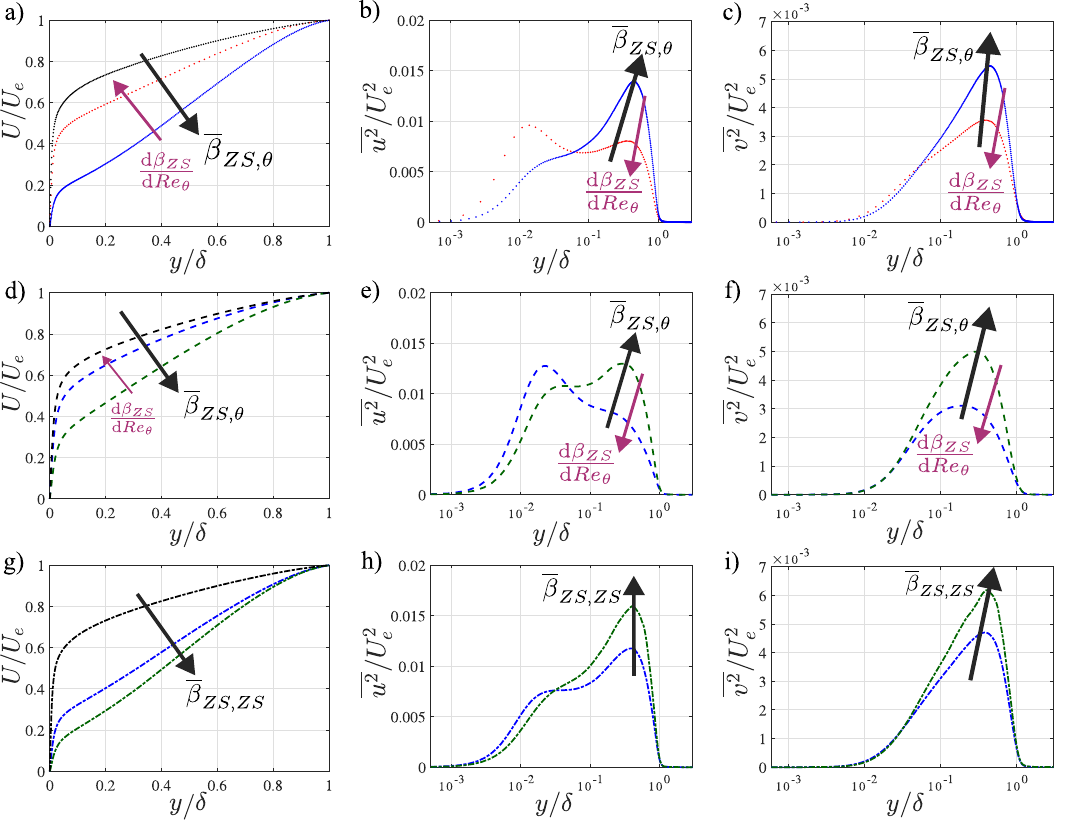}  
    \caption{Mean Statistics and Velocity variances for Cases 3 (top),4 (middle), and 5 (bottom). All quantities are scaled in outer units - $U_{e}$ and $\delta$. (a,d,g) Streamwise mean velocity, (b,e,h) $\overline{u^2}$ variances, (c,f,i) $\overline{v^2}$ variances. Colors and Line styles correspond to those in Tables \ref{tab1} and \ref{tab2}.}
    \label{fig5}
\end{figure*}

\subsection{TBLs imposed with strong adverse pressure gradients}

Next, we move to Cases 3 and 4 to study the effects of a prolonged APG history that lead to a large-defect TBL, by analyzing b1.4 and DNS22 at matched $\beta_{ZS}$ and $Re_{\theta}$. 
Since the friction velocity scale is not a suitable outer-velocity scale for large-defect cases, we selected the Zagarola-Smits velocity scaling to express both the local and historical pressure gradient parameters, with the latter computed based on $Re_{\theta}$ (see equation (\ref{eq2})).
The PG parameter $\beta_{ZS}$ has been shown previously to adequately represent the ratio of pressure force to turbulent force in the outer region of APG TBLs \citep{macielOuterScalesParameters2018}. On the other hand, it is worth noting that $Re_{\theta}$ is not logically compatible with $\beta_{ZS}$, as they do not originate from the same momentum equation analysis. 
In other words, they are not constructed using the same length and velocity scales. 
Nevertheless, since $Re_{\theta}$ is the most commonly used local Reynolds number for large defect TBLs, we test their combined use to assess whether they can account for flow history. Later in Case 5, however, we will examine the more consistent pair of $\beta_{ZS}$ and $Re_{ZS}$ and show consistency in the behaviours exhibited by Cases 3 to 5.

For Case 3, $Re_{\theta}$ and $\beta_{ZS}$ are matched, but the higher value of $\overline{\beta}_{ZS,\theta}$ in DNS22 suggests that it experiences a more pronounced historical effect of the APG than b1.4. Indeed, figure \ref{fig5}(a) shows the mean velocity defect increasing with $\overline{\beta}_{ZS,\theta}$, indicating a greater cumulative influence of the APG, consistent with our observations for $\overline{\beta}_{C,\theta}$ in Figures \ref{fig2}(top). 
Similarly, Figures \ref{fig5}(b) and \ref{fig5}(c) respectively show increasing outer peaks in the $\overline{u^2}$ and $\overline{v^2}$ variances with an increase in the $\overline{\beta}_{ZS,\theta}$ parameter. 
It is important to note here that the choice of an outer scaling-based PG parameter restricts our analysis solely to the outer region of the boundary layer, preventing reliable inferences about the inner region. 
We note that although the accumulated PG parameter $\overline{\beta}_{ZS,\theta}$ qualitatively aligns with the differences in the statistics, and is consistent with previously established trends of $\overline{\beta}_{C,\theta}$, the effectiveness of using $\overline{\beta}_{ZS,\theta}$ together with matched values of $Re_{\theta}$ and $\beta_{ZS}$ cannot be conclusively determined based on this case alone.

We thus consider Case 4 that compares DNS22 and DNS16 at matched $\beta_{ZS}$ and $Re_{\theta}$, enabling us to further assess the adequacy of using $\overline{\beta}_{ZS,\theta}$ (defined based on $Re_{\theta}$) for quantifying upstream PG history effects. 
Based on Figure \ref{fig5}(d), similar to Case 3, we observe an increasing mean velocity defect with $\overline{\beta}_{ZS,\theta}$. 
Cases 3 and 4 thus indicate that the parameter $\overline{\beta}_{ZS,\theta}$, when considered alongside matched values of $Re_{\theta}$ and $\beta_{ZS}$, may qualitatively represent the pressure force history and help explain the associated statistical trends in TBLs subjected to strong APG.
\cite{macielOuterScalesParameters2018} demonstrated that $\beta_{ZS}$ and $Re_{ZS}$ effectively follow the force balance in the outer region of APG TBLs, thus forming a more consistent set of parameters based on theoretical analysis.
Hence, we next consider Case 5, which compares DNS22 and DNS16 at matched $\beta_{ZS}$ and $Re_{ZS}$, with the accumulated PG parameter also computed based on $Re_{ZS}$ (see equation (\ref{eq3})). 
Figures \ref{fig5}(g-i) show that the mean velocity defect increases and velocity variances are more energized for DNS16 compared to DNS22, which aligns with the increasing $\overline{\beta}_{ZS,ZS}$ trend.
\begin{figure*}[t!]
     \centering
        \includegraphics[width=0.8\textwidth]{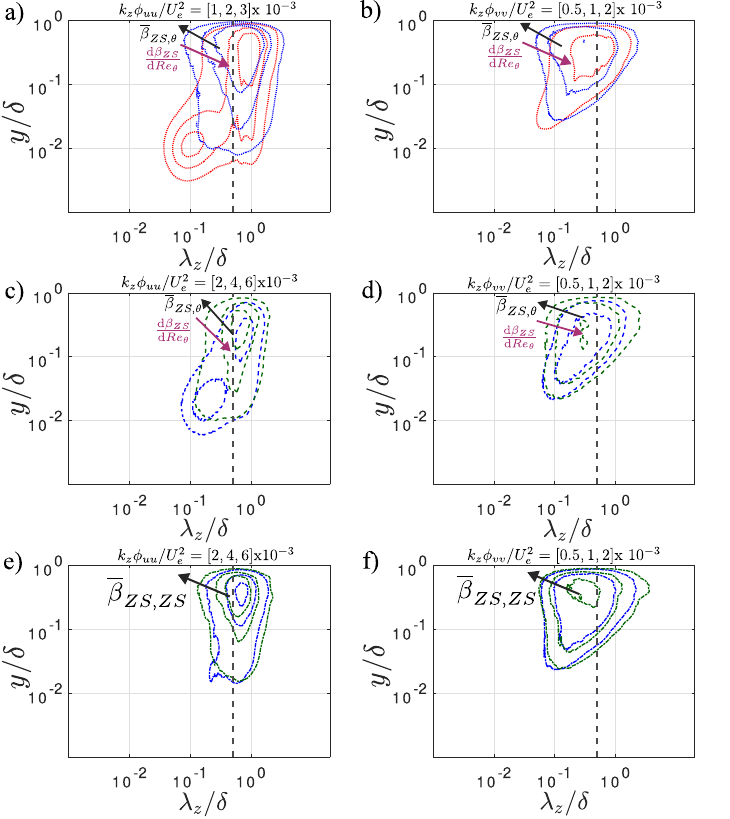}
        \caption{Contours of the premultiplied spanwise power-spectral density of streamwise and wall normal Reynolds stresses scaled in outer units ($k_z\phi_{uu}/U_{e}^2$,$k_z\phi_{vv}/U_{e}^2$) for Case 3 (a,b), Case 4(c,d), and Case 5(e,f). The dotted line corresponds to the $\lambda_{z,c}/\delta = 0.5$ scale cutoff computed based on \cite{deshpandeReynoldsnumberEffectsOuter2023d}. Colors and Line styles correspond to those in Tables \ref{tab1} and \ref{tab2}.}
        \label{fig6}
\end{figure*}

In terms of influence of local disequilibration, one can argue these effects are likely not to influence the statistics in Case 5 due to ${\rm d}{\beta}/{\rm d}{Re}$ values being of the same order of magnitude for the associated data sets (refer Table \ref{tab2}). 
For Cases 3 and 4, however, local disequilibration may play a more significant role, given that the ${\rm d}{\beta}/{\rm d}{Re}$ values differ by at least an order of magnitude between the two data sets being compared. Strong negative ${\rm d}{\beta_{ZS}}/{\rm d}{Re}_{\theta}$ for DNS22 (Case 3) and DNS16 (Case 4) imply a decreasing impact of the APG \citep{gungor2024b}, which acts in opposition to the strong positive value of  $\overline{\beta}_{ZS,\theta}$. 
The results for Cases 3 and 4 in Figure 4(a-f) indicate that the accumulated PG effect dominates, rather than the local disequilibration.

With the efficacy of $\overline{\beta}_{ZS,\theta}$ and $\overline{\beta}_{ZS,ZS}$ to quantify history effects demonstrated, next we leverage this new capability to understand the historical PG trends in power-spectral density and scale-decomposed contributions to velocity variances for these strong APG TBLs. 
\begin{figure*}[t!]
    \centering
    \includegraphics[width = \textwidth]{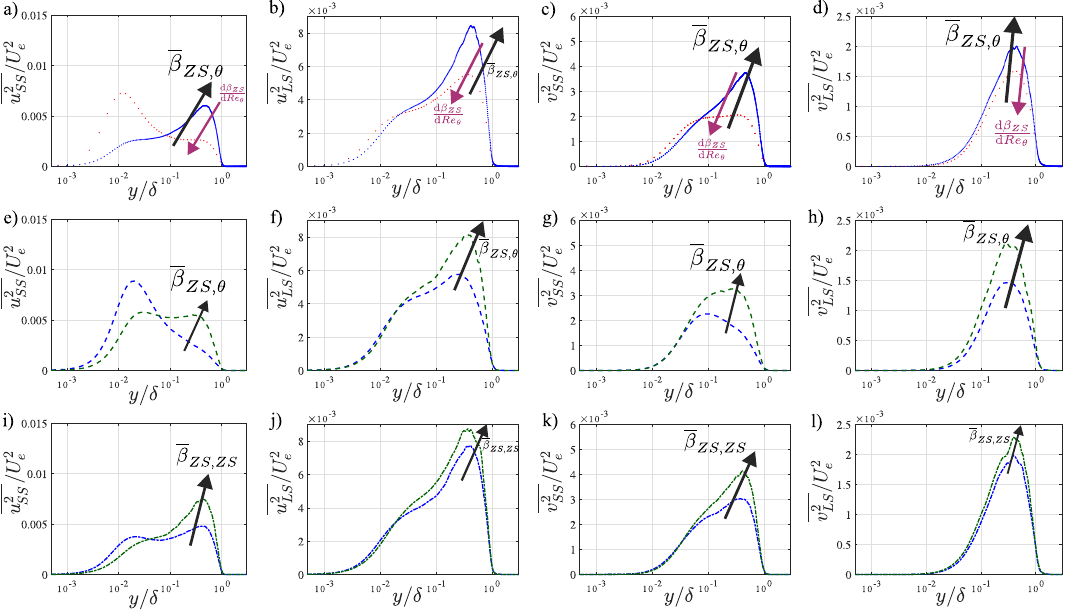}
    
    \caption{Small and large scale contributions to velocity variances in Cases 3 (top), 4 (middle), and 5 (bottom). Decomposition into small and large scales is done based on the $\lambda_{z,c}/\delta = 0.5$ cutoff discussed in $\S$\ref{subsec:analysis methodology}. All quantities are scaled with respect to outer units, $U_{e}$ and $\delta$. (a,e,i) Small scale contributions to $\overline{u^2}$ variances, (b,f,j) Large scale contributions to $\overline{u^2}$ variances, (c,g,k) Small scale contributions to $\overline{v^2}$ variances, (d,h,l) Large scale contributions to $\overline{v^2}$ variances. Colors and Line styles correspond to those in Tables \ref{tab1} and \ref{tab2}.}
    \label{fig7}
\end{figure*}
Figure \ref{fig6} presents premultiplied spanwise power-spectral density scaled in outer units for $\overline{u^2}$ and $\overline{v^2}$ in Cases 3, 4, and 5, complementing earlier inferences from velocity variances. The dotted line in Figure \ref{fig6} marks the scale cutoff ($\lambda_{z,c}/\delta = 0.5$, computed based on \cite{deshpandeReynoldsnumberEffectsOuter2023d}), to distinguish between small and large scales.
Figure \ref{fig6} for Cases 3, 4, and 5 shows similar trends of increasing outer-region energy with higher $\overline{\beta}_{ZS,\theta}$ and $\overline{\beta}_{ZS,ZS}$ values respectively, as observed in the low defect Case 2. In Cases 3 and 4, DNS22 and DNS16 displays outer-region small-scale energization and broader spectral spread, respectively, consistent with their higher $\overline{\beta}_{ZS,\theta}$, dominating the reduced APG effect caused by stronger negative disequilibration.

Case 5 reinforces this behavior - DNS16 exhibits increased energy at both ends of the spanwise wavelength spectrum for $\overline{u^2}$ and $\overline{v^2}$. Figure \ref{fig6}(e) shows a more energized and broader outer peak in $\overline{u^2}$ for DNS16, while Figure \ref{fig6}(f) captures a distinct outer peak in $\overline{v^2}$ that is absent in DNS22, both aligning with its greater $\overline{\beta}_{ZS,ZS}$ value.

Figure \ref{fig7} presents scale-decomposed contributions to $\overline{u^2}$ and $\overline{v^2}$ variances for Cases 3 to 5. 
We reiterate here that outer scaling limits our analysis of flow physics solely to the outer region. 
The plots reveal outer region energisation of both small- and large-scale contributions of $\overline{u^2}$ and $\overline{v^2}$ with increasing $\overline{\beta}_{ZS}$ (\ref{eq2} and \ref{eq3}), across all three high-defect Cases 3-5, consistent with Case 2 (Figure \ref{fig2} (below)).
Further, for Cases 3 and 4, the trend associated with the local disequilibration ($d{\beta}/d{Re}$; for both small and large-scales) is also consistent with that noted for Case 2 in the outer region.
\section{Summary and conclusions}
\label{sec:conclusion}

The present study utilizes four APG datasets - two highly resolved near-equilibrium low-$\beta$ LES datasets and two non-\\equilibrium DNS datasets reaching high values of $\beta$. 
The novelty of this study lies in the vast parametric range it spans in terms of APG ($\beta$) strength and Reynolds numbers ($Re$), allowing us to isolate the effects of upstream PG history and local disequilibration through identification of matched ($\beta$, $Re$, $\sim$ $d{\beta}/d{Re}$) and ($\beta$, $Re$, $\overline{\beta}$) cases, respectively.
We identified five matched cases across the four datasets using two PG parameters ($\beta_{ZS}$ and $\beta_C$) and three Reynolds numbers ($Re_{\tau}$, $Re_{\theta}$, and $Re_{ZS}$).
Here, the choice of the appropriate $\beta$ and $Re$ was motivated based on the strength of the APG TBLs under investigation. 
By construction, the accumulated PG parameter $\overline{\beta}$ does not account for the delayed response of the mean flow and turbulence, nor for the attenuation of pressure gradient effects with downstream distance. This study assesses the implications of this limitation on the ability of $\overline{\beta}$ to represent pressure gradient history effects.

Case 1, which represents a previously established comparison of low-defect TBLs at matched $\beta_C$ and $Re_{\tau}$, lays the foundation for comparing statistical trends in the subsequent cases. Though $\overline{\beta}_{C,\theta}$ can qualitatively represent the flow history and indicate trend differences, the parameter constructed using $Re_{\theta}$ is inconsistent with the choice of matched $Re_{\tau}$. This prompted our switch to matched cases of $\beta_C$ and $Re_{\theta}$, which is consistent with the definition of $\overline{\beta}_{C,\theta}$.

In addition to fortifying our inferences on the effects of accumulated PG on TBL statistics, Case 2 offers insight into the role of local disequilibration, which leads to a delayed turbulence response.
The rapid increase in the APG strength (${\rm d}{\beta_{C}}/{\rm d}{Re_{\theta}}$) in DNS22 manifests into a delayed turbulence response resulting in statistics with lower energisation than expected for the local $\beta$ value, particularly affecting the outer region of the TBL. 
Case 2 thus highlights the limitation of the accumulated PG parameter in capturing history effects in situations involving rapid disequilibration, which induces a pronounced delayed response in turbulence. Case DISEQ further highlights the impact of local disequilibration on turbulent scales, revealing that the near-wall small-scale region adapts rapidly to the changing pressure gradient, while other regions exhibit a delayed response.

Since Cases 3 to 5 are high-defect cases, pressure gradient parameters based on Zagarola-Smits scales were adopted. 
This was inspired from the findings of \cite{macielOuterScalesParameters2018} who showed that $\beta_{ZS}$, unlike $\beta_C$, represents the ratio of pressure force to turbulent force across all APG conditions. The comparison of TBLs at matched values of $Re_{\theta}$ and $\beta_{ZS}$ (Cases 3 and 4) revealed qualitative consistency with the upstream influence of the APG, as expressed by $\overline{\beta}_{ZS,\theta}$, thereby suggesting that $\overline{\beta}_{ZS,\theta}$ may qualitatively represent the pressure force history.
Case 5 compared DNS22 and DNS16 at matched $\beta_{ZS}$ and $Re_{ZS}$, representing parameters logically compatible based on the momentum equation. 
The accumulated PG parameter $\overline{\beta}_{ZS,ZS}$, derived from these two parameters, was qualitatively consistent with the observed trends.
The present study thus establishes that appropriate scaling based on APG strength enables, at least qualitatively, the upstream pressure gradient history to be represented using the accumulated PG parameter, regardless of the APG TBL strength (i.e., from weak to strong).

According to \cite{bobkeHistoryEffectsEquilibrium2017}, \citet{montyParametricStudyAdverse2011} and \citet{tanarroEffectAdversePressure2020a}, small- and large-scale structures are energized with an increase in $\beta$ or $\overline{\beta}$. 
This observation, previously limited to low-defect cases, can now be extended to a broader range of APG strengths with the establishment of the $\overline{\beta}_{ZS}$ parameter.
The present analysis confirms the turbulence energisation of both small- and large-scales in the TBL outer region, with increasing APG history effects for both low- and high-defect cases. 
This study can be viewed as one of the first steps towards proposing a unifying methodology for characterizing history and local disequilibration effects on turbulent boundary layers for varying PG strengths/defects (as witnessed in various practical applications), through consideration of classical scaling arguments.
In that sense, the present study lays the groundwork for future studies aiming to delve deeper into the parametric characterisation of TBLs with varying APG strengths and upstream PG histories.
An example is the past work of \cite{vinuesaRevisitingHistoryEffects2017} that modeled $C_f$ and $H_{12}$ correlations using ZPG data and the accumulated PG parameter $\overline{\beta}_{C,\theta}$, but was limited to low-to-moderate $\beta_C$ TBLs.
Future studies can focus on extending this analytical modeling to high-defect TBLs with establishment of the $\overline{\beta}_{ZS}$ parameter in the present study. 
However, it remains to be determined whether a more sophisticated accumulated PG parameter is required, as the current one is found not to account for the delayed flow response to PG disequilibration, or for the attenuation of PG effects with upstream distance.

\section*{Data availability}
\noindent
All the databases employed in this manuscript are publicly available at the following links:
\underline{ZPG, b1.4 and b1.0}: \href{https://www.vinuesalab.com/apg/}{weblink1} \\
\underline{DNS22 and DNS16}: \href{https://yvanmaciel.gmc.ulaval.ca/databases/}{weblink2}

\section*{Code availability}
\noindent
The code used for analysis can be requested after publication.

\section*{Competing interests}
\noindent
The authors declare that they have no competing interests. 

\section*{Acknowledgements}
\noindent
The authors are grateful to Dr R. Pozuelo for sharing his data set.
RD is grateful to the University of Melbourne's Postdoctoral Fellowship for financial support.
RV acknowledges the financial support from ERC grant no. ‘2021-CoG-101043998, DEEPCONTROL’. Views and opinions expressed are however those of the authors only and do not necessarily reflect those of the European Union or the European Research Council. Neither the European Union nor the granting authority can be held responsible for them.
TRG and YM acknowledge PRACE for awarding us access to Marconi100 at CINECA, Italy and Calcul Québec (www.calculquebec.ca) and the Digital Research Alliance of Canada (alliancecan.ca) for awarding us access to Niagara HPC server. TRG and YM acknowledge the support of the Natural Sciences and Engineering Research Council of Canada (NSERC), project number RGPIN-2019-04194. TRG was supported by the research fund of Istanbul Technical University (project number: MDK-2018-41689)
\section*{Author contributions}
\noindent
\textbf{AM}: Conceptualization, Methodology, Analysis, Writing - Original draft. 
\textbf{RD}: Conceptualization, Methodology, Supervision, Writing - Review \& Editing.
\textbf{TRG}: Analysis, Writing - Review \& Editing.
\textbf{YM}: Methodology, Supervision, Writing - Review \& Editing.
\textbf{RV}: Conceptualization, Supervision, Writing - Review \& Editing, Funding acquisition.

    \appendix
    \section*{Appendix A: Validation of outer-scaled spanwise wavelength cutoff}
    \label{sec:appendix_A}
\begin{figure*}[t!]
    \centering    
    \includegraphics[width=\textwidth]{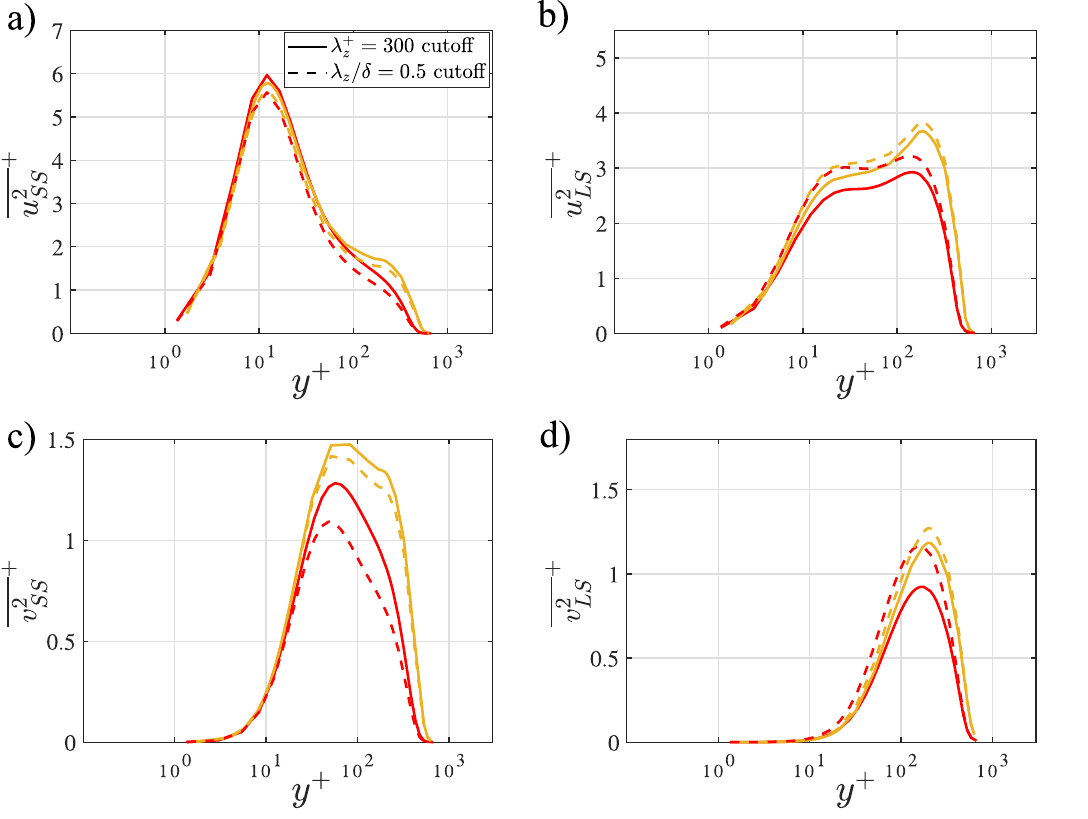}
    \captionof{figure}{Scale-decomposed $\overline{u^2}$ and $\overline{v^2}$ variances for Case 1 decomposed based on a viscous-scaled ($\lambda^+_{z,c}$=300) and outer-scaled spanwise wavelength cutoff ($\lambda_{z,c}/\delta$ = 0.5). Contributions evaluated using viscous and outer-scaled cutoff are indicated by solid and dotted lines, respectively. Colors correspond to datasets mentioned in Table \ref{tab1}. }
    \label{fig8}
\end{figure*}

    This appendix section serves to justify the use of the outer-scaled spanwise wavelength cutoff $\lambda_{z,c}/\delta = 0.5$ to decompose $\overline{u^2}$ and $\overline{v^2}$ variances into their large and small-scale counterparts for large-defect cases. 
    We validate the choice by comparing the scale-decomposed variance plots computed based on the outer-scaled cutoff for Case 1, with those computed using the $\lambda_{z,c}^+ = 300$ cutoff proposed by \cite{deshpandeReynoldsnumberEffectsOuter2023d}. 
    Figure \ref{fig8} reveals a reasonable agreement between the scale-decomposed variances computed using the two cutoffs, thus confirming the outer scaled cutoff as a sensible choice.

\appendix
    \section*{Appendix B: Validating the choice of Quasi-matched PG strength and Reynolds numbers}
    \label{sec:appendix_B}
    
\begin{figure*}[t!]
    \centering   
    \includegraphics[width=\textwidth]{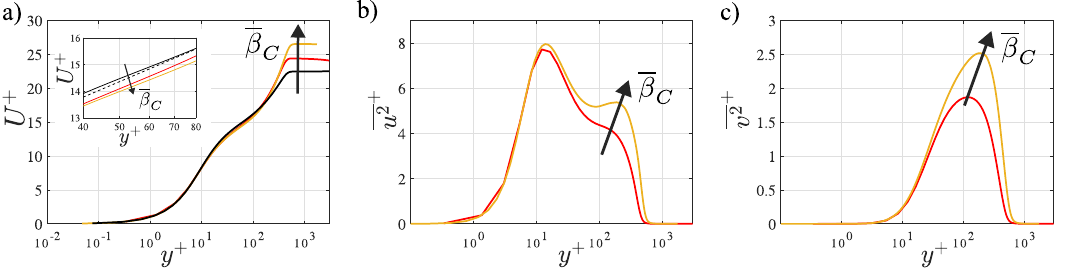}
    \captionof{figure}{Mean streamwise velocity profiles, $\overline{u^2}$ and $\overline{v^2}$ variances for Quasi-Matched parameters corresponding to Case 1. All quantities are scaled in viscous units - $U_{\tau}$ and $l_{\tau}$. (a) Streamwise mean velocity, (b) $\overline{u^2}$ variances, (c) $\overline{v^2}$ variances. Colors and Line styles correspond to datasets mentioned in Tables \ref{tab1} and \ref{tab2}.}
    \label{fig9}
    \end{figure*}

    This section of the appendix is presented to reaffirm that the variation of turbulence statistics corresponding to "Quasi-matched" positions (see Table \ref{tab2}) are consistent with those noted for the Exactly-Matched cases. 
    To demonstrate this, we compare mean streamwise velocity, $\overline{u^2}$ and $\overline{v^2}$ variances for the Quasi-Matched Case 1 in Figure \ref{fig9}, the general trends inferred from which remain consistent with those noted from the Exactly-Matched Case 1 in Figure \ref{fig3} (top). 
    Similar consistency in trends is maintained between all other Exactly-Matched and Quasi-Matched Cases 2-5 in noted in Table \ref{tab2}.

\bibliographystyle{elsarticle-num-names} 
\bibliography{refs}

\end{document}